\documentclass[conference]{IEEEtran}
\usepackage{cite}
\usepackage[pdftex]{graphicx}
\usepackage{amsmath}
\usepackage{algorithmic}
\usepackage{array}
\usepackage{caption} 
\usepackage{amsmath}
\newtheorem{theorem}{Theorem}

\newtheorem{claim}{Claim}

\begin{document}
\IEEEoverridecommandlockouts

\title{Fast direct access to variable length codes}
\author{\IEEEauthorblockN{Boris Ryabko}
\IEEEauthorblockA{Federal Research Center for Information and Computational Technologies
 of SB RAS\\ 
Novosibirsk state university
\\Novosibirsk, Russian Federation \\Email: boris@ryabko.net\\}
}

\maketitle

\begin{abstract}
We consider the issue of direct access to any letter of a sequence encoded with a variable length code and stored in the computer's memory, which is a special case of the random access problem to compressed memory. The characteristics according to which methods are evaluated are the access time to one letter and the memory used. The proposed methods, with various trade-offs between the characteristics, outperform the known ones.
\end{abstract}

\textbf{keywords:}  source coding, data compression, random access, direct access, compressed memory.

\section{Introduction and a statement of the problems}
Variable-length codes are widely used for transmitting and storing data, because they can significantly reduce their size. In applications, the following model is usually used: there is a sequence of letters $x =  x_1 x_2 .... x_N $, and these letters are encoded by a lossless code with codewords  of different lengths. In the case of data  transmission,  the decoder decodes the letters sequentially, and the difference in length is not a problem, but in many storage applications it is necessary to  access directly  some letters $ x_i $ in the sequence. That is, the compressed data must be available in  random order. This is not an easy task, since the place of the codeword $ x_i $ depends on the total length of the codewords of the previous letters $ x_1 ... x_ {i-1} $ and in this case the decoder must  calculate this  total length, which can take time.   
So, the implementation of direct access requires an additional consideration.  This problem was investigated in numerous papers (see \cite{directec,ku,directec-modern,squ,bm-,ne} and reviews therein).

The method of direct (or random) access is the basis of some other operations that are used to work with compressed data. Among these operations, the following two are considered primitives for many others: the $ rank_a (i) $ operation returns the number of occurrences of a letter $ a $ up to $ i $ in the compressed sequence, and $ select_a (i) $ returns the position of the $ i $th occurrence of $ a $, see \cite{directec,ku,directec-modern,squ,bm-,ne,fm,ks,20,bil}.
\cite {squ,bm-} shows that, given a data structure that only supports direct access, it is possible to build an efficient method for $ rank_a (i) $ and $ select_a (i) $.
Therefore, we focus on the problem of direct access.
Also note that, as is customary in this area, we assume that all operations are performed on a random access machine (RAM), which is a model of a ``regular'' computer.

Let us consider an example.
Suppose that letters of the encoded sequence $x$ belongs to an alphabet $A = \{a_0, a_1, ..., a_{L-1}\}$, where $L \ge 2$. Let    $bin_m (n)$
be the binary representation of $n$ presented as $m$-bit word. For example, $bin_4(2) =
0010$ (here and below  $\log \equiv \log_2$). 
Possibly, the simplest   code of the letters from $A$ is given by the following equation:
 \begin{equation}\label{simp}
code (a_i) = bin_l(i) \, ,
 \end{equation}  
 where $l = \lceil \log L \rceil $.
 Here the time of decoding of any letter $x_i$ from $x=x_1 $ $ x_2 .... x_N $ is $O(1) $ and the size of the encoded $x$ equals $N \,  \lceil \log L \rceil $.
  
If the probabilities $ p (a) $ of  different letters from $A$ are not equal, the size of the encoded sequence can be reduced using variable length codes. This problem is well known in information theory \cite{co} and, in particular, it is known that the Huffman code $ C_H $ has the minimum average length $ E (C_H) $, for which
   \begin {equation} \label{haf}
h_1 (p) \le E (C_H) <h_1 (p) +1 \,
   \end{equation}
   where $ h_1 (p) = - \sum_{i = 0}^{L-1} p (a_i) \log p (a_i) $ is the first-order Shannon entropy \cite{co},
   $ E (C_H) =  \sum_ {i = 0}^{L-1} p (a_i) | C_H (a_i) | $ is the expectation, $ C_H (a) $ is the codeword of the letter $ a $ and $ | v | $ is the length of $ v $, if $ v $ is a word, and the number of elements, if $ v $ is a set.
   
   For example, let $p(a_0) = 1/2,  p(a_1) = 1/4, $ $..., $ $p(a_{L-2} )= 1/ 2^{L-1}$ and $ p(a_{L-1} )= 1/ 2^{L-1}$.  
If we apply the Huffman code, then   $|C_H(a_0) | = 1,    |C_H(a_1) | = 2, ...,  $ 
$|C_H(a_{L-2}) | = L-1$  and
$ |C_H(a_{L-1}) | = L-1$ 
(see \cite{co}), and, hence, $E(C_H) = h(p) = 
\sum_{i = 0}^{L-2} 2^{-(i+1)} (i+1) \, + 2^{-(L-1)} (L-1)$ $<2$.
In this case, the size is  about $ 2 N $, which is much less than $ N \, \lceil \log L \rceil $ of the first method for large $ L $, but the time to decode one letter is proportional to $ N $ if we decode letters sequentially (letter by letter), that is, much more than in the first case.


So, we see that there is a certain trade-off between the size of the encoded data and the speed of access to their parts.
A fairly complete overview of known methods is given in \cite{directec,directec-modern}, which describes several methods with varying trade-offs.
In this article, we describe several new compact storage methods with varying trade-offs between the size of the stored data and the access speed, showing that some methods are close to optimal and outperform known methods.

The rest of the article consists of two parts.  In the first part, we consider the random access problem for the case where encoding and decoding are performed using separate letters, while the other part describes generalizations in which a sequence is divided into subsequences that are encoded separately or combined into blocks, where the blocks are treated as new letters.

\section{General method of fast random access}\label{p1} 
In this part   we consider what is perhaps, the simplest situation, where there is an $ N $ -lettered sequence from the $ L $ -lettered alphabet $ A $ and the frequency of occurrence of all $ a \in A $ are known.  The goal is to build a storage system for this sequence.
(A close mathematical model is as follows: 
there is a  source that generates letters from the $ L $ -lettered alphabet $ A $ with known probabilities of the letters, and one needs  to build a storage system for the $ N $ -lettered sequence generated by this source.) 

\subsection{Trimmed codes } 

Let, as before,  there be an alphabet $A=\{a_0, ... , $ $ a_{L-1} \}$, $L\ge 2$, with a probability distribution
$p = p(a_0), ... ,$ $ p(a_{L-1}) $ and $C$ be a   lossless code for letters from $A$.  
In the previous Huffman code example, we saw that the length of the codeword can range from 1 to $L-1$.
In the following applications it will be convenient to use  codes  for  which the code length of any letter is not greater than $\lceil \log L \rceil +1$.  (Note that for any code the maximal length of the codewords is not less than $\lceil \log L \rceil $.) We call such codes as trimmed and define one of  them as follows:
 if $C$ is a code then  
\begin {equation}\label{tr}
C^{tr}(a_i) =
  \begin{cases}
    0\, C(a_i)       & \quad \text{if } |C(a_i) |\le \lceil \log L \rceil \\
    1\,  bin_{\lceil \log L \rceil}(i) & \quad \text{if }   |C(a_i) | > \lceil \log L \rceil  \, .
  \end{cases}
\end{equation}
Let us explain how to decode. First,  the decoder reads the first binary letter. If it is $0$, the decoder uses the codeword of the code $C$ in order to find the encoded letter. If the first letter $1$, the next $\lceil \log L \rceil$ letters contain the binary notations of $i$, i.e. the letter is $a_i$.  Clearly, $|C^{tr}(a) | \le  \min \{\lceil \log L \rceil +1, $ $|C(a) | +1 \}$ 
for any letter $a$. 

 Note, that  from this   we can estimate the average length (per letter) of the trimmed  code:   
\begin {equation}\label{haf-tr} 
 E (|C^{tr}|) <E(|C|)+1 \, ,
\end{equation} 
where $E(\,)$ is the expectation.
So, this design reduces the length of long codewords, but one extra bit is charged for this reduction.

Let us consider an example of the trimmed Huffman code.
Let there be an alphabet $A= \{a_0, ... , a_{14} \}$ with probabilities $p(a_0) =2^{-1},  p(a_1) =2^{-2}, p(a_2) =2^{-3}, ..., p(a_{13}) =2^{-14}, p(a_{14}) =2^{-14}$. Clear, $C_H(a_0) = 0$, 
$C_H(a_1 ) = 1 0, $ $C_H(a_2) = 110, $
$C_H(a_{13}) = 11111111111110 $ (thirteen 1s and 0), $C_H(a_{14}) = 11111111111111 $ (fourteen 1s). Taking into account that $\lceil \log 15 \rceil  = 4$  and the definition (\ref{tr}) we obtain
\begin {equation}\label{haf-tr1}
C_H^{tr}(a_0) = 0 \,0, C_H^{tr}(a_1) = 0 \,10, C_H^{tr}(a_2) = 0 \,110, 
$$ $$
C_H^{tr}(a_3) = 0 \,1110,  
 ... , C_H^{tr}(a_{13}) = 1 \,1101, C_H^{tr}(a_{14}) = 1 \,1110 \, .
\end{equation} 
So, we can see that the long code-words became shorter, but the length of short codewords is increased by 1.

In this report, we consider perhaps the simplest trimmed code (\ref{tr}), but there are more efficient codes of this kind. For example, there are methods for constructing a code with a minimum average codeword length, provided that the maximum codeword length is limited by some constant $ \Delta $, see  \cite{limh,limh2}
(of course, $ \Delta $ $\ge  \lceil \log L \rceil$).
Thus, the average length of the codeword of the trimmed code constructed by these methods for $ \Delta = \lceil \log L \rceil + 1 $ may be less than for the described code $ C_H^{tr} $.
Besides, there are other possibilities to construct trimmed codes, but, anyway,  the benefit will be not greater than 1 bit, if we take into account that the Huffman code is optimal and   (\ref{haf-tr}).

It is worth noting that there are fast decoding codes \cite{rae,rr,fastd} that can come in handy for storing data.

\subsection{The storage system based on the trimmed   codes}

Here we describe a system whose decoding time is $ O( \log N)$ (instead of $ const  \, N $ for Huffman code), but 
the required memory (per letter) is $ h_1(p) + \log \log L + O(1)$ instead of $ h_1 (p) $ for Huffman code, where $ h_1 (p) $ is the Shannon entropy. We do this in the next two small sections.

\subsubsection{The preliminary  version of the storage system}

The following storage system $\Sigma^{tr}_H$ consists of two codewords $ y $ and $ z $:
\begin {equation}\label{yz} 
y =  y_1 y_2 ... y_N  \, \text{and } z= z_1 z_2 ... z_N \, ,  \text{where }  z_i = C_H^{tr}(x_i),
$$
$$\, y_i = bin_{\lceil \log \lambda  \rceil} | C_H^{tr}(x_i)|\,  
, \lambda = \lceil \log L \rceil  +1,  i=1, ..., N.
\end{equation} 
Taking into account (\ref{haf}) and (\ref{haf-tr}), from this definition we obtain
$|z| \le N( h_1(p) +2)$,  $|y| = N \lceil   \log \lambda  \rceil $ $\le N ( \log (\log L +2)$ $ +1)$, and, hence, 
\begin {equation}\label{length-yz} 
|x| + |y| \le N  (h_1(p) +   \log ( \log L  +2)    + 3)
\end{equation} 
Suppose we want to find the letter $ x_i $, that is, find such $ a_j \in A $ that $ x_i = a_j$.
From the definitions (\ref{yz}) and (\ref{tr}) we can see that any letter $ x_i $ can be obtained from $ z $ by sequentially decoding the letters $ x_1 ... x_i $. 
It is clear that the average decoding time is proportional to $ |z| = N  h(p) $.
The point is that the use of additional $ y $ makes it possible to reduce the decoding time to  $ N \lceil  \log \lceil \log    (L+1)  \rceil  \rceil $ .
 Indeed, if we want to find the letter $ x_i $, we can read the letters $ y_1 ... y_ {i-1} $ and calculate the sum of the integers represented there, that is, $ \sum_ {k = 1}^{i- 1} y_k $.
It is important that this sum equals the sum of the codewords length of $x_1 ... x_{i-1}$, that is,  $ \sum_ {k = 1}^{i- 1} y_k $ $=  \sum_ {k = 1}^{i- 1} |C^{tr}_H(x_k)| $, see
(\ref{yz}). So, the first bit of the word $C^{tr}_H$ in $z$ is   $ \sum_ {k = 1}^{i- 1} y_k +1$ and the length of this word is written in $y_i$. 

The described properties of the  storage system $\Sigma^{tr}_H$ are summarised  in the following 
\begin {claim}\label {c1}
Let there be a source generating letters from the alphabet $ A = \{a_0, ..., a_{L-1} \} $ with the probability distribution $ p $ and $ x = x_1 ... x_N $ be  the sequence generated by this source. If the storage system  $\Sigma^{tr}_H$ is applied  to  sequence $x$
the average  time of  decoding of one letter equals the time of calculation of the sum 
$ \sum_ {k = 1}^{i- 1} y_k $ and the average memory space is $|z| + |y|  \le N  (h_1(p) +   \log ( \log L  +2)    + 3)$ bits.
\end{claim}
Let us consider an example. Let the alphabet and the truncated Huffman code be as in the previous example (\ref{haf-tr1}) and the generated sequence $x = x_1 ... x_4$ be $a_5 a_0 a_0 a_{10}$.
Then
$$ z = 10101\, 00\,00\, 11010 , \,\, y = 101\,010\,010\,101 \, .
$$

\subsubsection{Acceleration with a binary indexed tree}
Here we describe a new storage system using the so-called binary indexed tree \cite{r1, r2} to speed up the the calculation of sums $ \sum_ {k = 1}^{i- 1} y_k $  for different $i$. We denote this storage system  by $\Sigma^{bit}$.
Let's describe a simplified binary index tree for our purposes. To simplify the notation, we assume that $ N = 2^\nu $, where $\nu$ is an integer,  but the generalization to an arbitrary $ N $ is straightforward. 

Define the auxiliary values $Q$  as follows
\begin{equation}\label{qq}
Q^1_1 = y_1,  Q^1_2 = y_2, \, ... , Q^1_{N}  = y_{N} , \qquad 
$$
$$ 
Q^2_1 = y_1+y_2, Q^2_2 = y_3+y_4, ..., 
Q^2_{N/2} = y_{N-1}+y_{N} ,
$$ $$
Q^3_1 = y_1 +... +y_4, Q^3_2 = y_5+ ... +y_8, .... , Q^3_{N/4}= y_{N-3} + ... +y_N
$$ 
$$,  ... , \qquad  \qquad  \qquad  \qquad  \qquad  \qquad  \qquad  \qquad \qquad  \qquad  \qquad  \qquad$$
$$
Q^{\log N -1}_1 = y_1+ ... +y_{N/2 },  \, \, Q^{\log N -1}_2 = y_{N/2 \, +1}+ ... +y_{N},
$$ $$
Q^{\log N }_1 = y_1+ ... +y_{N } \, .
\end{equation} 
Then, remove all even $Q$s and obtain the following set $\hat{Q}$:
\begin{equation}\label{qqh}
\hat{Q}^1_1 = y_1,  \hat{Q}^1_3 = y_3, \, ... ,  \,\, \hat{ Q}^1_{N-1}= y_{N-1} , \qquad 
$$
$$ 
\hat{Q}^2_1 = y_1+y_2, \hat{Q}^2_3 = y_5+y_6, ...,   \hat{Q}^2_{N/2 -1} = y_{N-3}+y_{N-2} ,
$$ $$
\hat{Q}^3_1 = y_1 +... +y_4, \hat{Q}^3_3 = y_9+ ... +y_{12}, .... , $$ $$ \hat{Q}^3_{N/4-1}= y_{N-7} + ... +y_{N-4} $$ $$
\, ,  ... ,  \quad 
\hat{Q}^{\log N }_1 = y_1+ ... +y_{N }   \, \, .  \qquad   \qquad  \qquad  \qquad  \qquad 
\end{equation} 
To shorten the notation, define $ \sigma = \lceil  \log  \lceil \log ( L+1)   \rceil    \rceil $.
Given that any $ y_i $ is an integer stored in a $ \sigma $-bit words, we will store the sum of $ \hat{Q}^2_i $ pairs in a $ (\sigma + 1) $-bit words, $ \hat{ Q}^3_i $ in $ (\sigma + 2)$- bit words, $ ... ,$ $ \hat{Q}^k_i $ in $ (\sigma + k - 1) $-bit words. 
 So, the total memory size $M$ for 
$\hat{Q}^k_i$, $k = 1, 2, ..., \log N -1$, $i = 1, 3,5, ... , N/2^k - 1$ is as follows:
$$ M =  \sigma N/2 + (\sigma +1) N/4 +  (\sigma +2) N/8  +  (\sigma +3) N/16 + ...  \, .
$$ 
From this we can derive the following estimate: 
$$ M =  N ( \sigma   \, \, (1/2+ 1/4 + 1/8 + ... ) \,\,\, +  ( 1/4 + 2/8 +3/16 + ... ) = 
$$
$$ N (\sigma  + \frac{1}{4} (1 + \frac{2}{2} + \frac{3}{4} +    \frac{4}{8}\  + \frac{5}{16} + ...) )=  N (\sigma +1) \, .
$$ (Here we used the identities $\sum_{i=0}^\infty p^i = 1/(1-p)$ and  $\sum_{i=0}^\infty  $ $ i p^{i-1} = $ $1/(1-p)^2 $ for $p = 1/2$).

The set $ \hat {Q} $ is stored instead of the sequence $y$. 
So, if we compare the amount of memory required for this method and the previous one, we can see that the difference is 1 bit per letter (i.e. $ N (\sigma +1) $ instead of $ N \sigma $).

The most time-consuming part of the $\Sigma^{tr}_H$ storage system is the calculation of the sums $\sum_{k=1}^j y_k$. The set $ \hat {Q} $  is a tool to speed up this computation, and we now show how $ \hat {Q} $ makes it possible to compute these sums faster.

First, we consider an informal example in order to  to explain the main idea. 
Let $N= 8$ $(\nu =3)$. Then 
$$
\hat{Q}^1_1 = y_1,  \hat{Q}^1_3 = y_3, \,  \hat{ Q}^1_{5}= y_{5},  \,\, \hat{ Q}^1_{7}= y_{7} , \qquad 
$$
$$ 
\hat{Q}^2_1 = y_1+y_2, \hat{Q}^2_3 = y_5+y_6, 
$$ $$
\hat{Q}^3_1 = y_1 +... +y_4,  \, \hat{Q}^4_1 = y_1 +... +y_8
$$
Then, $ y_1+ ...+ y_7$ can be calculated as $\hat{Q}^3_1 +  \hat{Q}^2_3 + \hat{ Q}^1_{7}$ (that is, $ y_1+ ...+ y_7$ $= ( y_1+ ... +y_4)$ $+ ( y_5 +y_6) +(y_7)$).
Analogically, $ y_1 +... +y_5$ $= \hat{Q}^3_1 + \hat{Q}^1_5 $,  $ y_1+ ... +y_3$
$= \hat{Q}^2_1 + \hat{Q}^1_3 $, $ y_1+ ...+ $ $y_8$ $= \hat{Q}^4_1$,  etc. 

The formal description of the calculation    $\sum_{k=1}^j y_k$ is as follows:
First, we  present  $j $ in the binary system 
$bin_{(\nu +1)}( j) =  (\alpha_{0} \alpha_{1} ...   \alpha_{\nu})$.
Then corresponding $\hat{Q}$ are summarised in $\nu$ steps as follows:

{\bf Define two integers $S$ and $T$ and let the algorithm be as follows:
$$S:=0, T:=0  \, , $$
For \,\,$ i =0, ... , \nu  \,\, $do 
$
\{ t := 2T+ \alpha_i,  \, \quad
S:= S + \alpha_i  \hat{Q}_T^{\nu - i +1} \}, 
$
$$
 \sum_{k=1}^j y_k := S  \, .$$ 
}

If we apply this algorithm to the previous example and calculate $\sum_{k=1}^5 y_k$, we obtain
$$bin_4(5) = (0 1 0 1),  S=0, T = 0; \, i = 1, T = 0+1, S= 0+ 1  \hat{Q}_1^{3}; $$
$$
i=2, T= 2+0, S = \hat{Q}_1^{3} +0 ;  i=3, T= 2\times 2 +1, S = \hat{Q}_1^{3} + \hat{Q}_5^{1};
$$ 
Finally, $\sum_{k=1}^5 y_k$ $=\hat{Q}_1^{3} + \hat{Q}_5^{1}$.

Let us estimate the time of calculation of the described algorithm. 
There are $\nu $ steps where several operations of summation are carried out.
So, the number of operations is proportional to $ \nu $. The length of all integers is $ O (\nu +\sigma) $ and hence the time (in bit operations) is proportional to $ \nu (\nu + \sigma)$ $= \log N (\log N + \log \log L)$.

Thus, we have estimated the memory space and the time of the decoding, and can summarise 
   the properties of the storage system $\Sigma^{bit}$   as follows:
\begin {theorem}\label {c1}
Let there be a source generating letters from the alphabet $ A = \{a_0, ..., a_{L-1} \} $ with the probability distribution $ p $, and let $ x = x_1 ... x_N $ be  the sequence generated by this source. If the storage system  $\Sigma^{bit}$ is applied  to a sequence $x$, then
the average  time of  decoding of one letter is proportional  $\log N (\log N + \log \log L)$ and the average memory space is not grater than  $  N  (h_1(p) +   \log \log (L+2)    + 4)$ bits.
\end{theorem}

\section{Variants of the proposed method that yield different time-memory trade-offs.}
In this part, we will describe some versions of the proposed method  $\Sigma^{bit}$ that are designed to represent various trade-offs between time and memory. Some of them are superior to known random access algorithms from  \cite{directec,ku,directec-modern}.

Let us first consider a popular storage scheme for compressed data, where the sequence $ x = x_1 ... x_N $ is represented as $ m $ subsequences $ x ^ 1 = x_1 ... x_M $, $ x ^ 2 = x_ {M + 1} ... x_ {2M} $, $ ... $, $ x ^ m = x _ {(m-1) M + 1} ... x_N $ and $ N = m M $.
To simplify the notations, 
  suppose that $ N$ and $M$ are a power of two.
The storage system $ \Sigma^{bit} $ is then applied to any subsequence of $ x^i $ separately, but the starting address of any encoded sequence is stored in computer memory (this requires $ m\log( N  \lceil \log L  \rceil ) $ bits). 
If someone wants to find the letter $ x_i $ from $ x $, he must first calculate $ r = \lceil  i / M   \rceil $ and $ j = i - ( r-1) M $ and then find the $ j $th letter in $ x^{ r } $ by the described decoding method. The time of the first part is proportional $\log N$ and, 
hence,  the extra time per letter is $ \log N \, $. 
Denote this system as 
$ \Sigma^{N,m}$. So, from this description and from Theorem 1 we obtain the following
\begin {theorem}\label {c2}
Let there be a source generating letters from the alphabet $ A = \{a_0, ..., a_{L-1} \} $ with the probability distribution $ p $, and let $ x = x_1 ... x_N $ be  the sequence generated by this source. If the storage system  $\Sigma^{N,m}$ is applied  to the sequence $x$, then
the average  time  of  decoding of one letter ($T$) is proportional to $\log M (\log M + \log \log L) + $  $ \log N $ and the  memory space ($S$) is not grater than $  N  (h_1(p) +  \log \log (L +2)   + 4)+ $ $ m\log( N  ( \log L  +1) $ bits, where $M = N/m$.
\end{theorem}

Let us consider  an example. 
Suppose, 
$M= N^\alpha, \alpha \in (0,1)$. 
Then, asymptotically, for $N \to \infty$, $\quad$
$S= N(h_1 + \log \log L + O(1) ) $ and  $T$ is proportional to $\alpha^2 (\log N)^2 \, (1+o(1)) $.
So, we can see that the time of encoding is significantly less than for the initial method $\Sigma^{bit}$, whereas the memory size is asymptotically the same.

Another popular storage scheme for compressed data  $ \hat{\Sigma}^{N,m}  $is as follows: Again, 
the sequence $ x = x_1 ... x_N $ is represented as $ m $ subsequences $ x ^ 1 = x_1 ... x_M $, $ x ^ 2 = x_ {M + 1} ... x_ {2M} $, $ ... $, $ x ^ m = x _ {(m-1) M + 1} ... x_N $, $ N = m M $, but now any subsequence $x^i$ is considered as a letter from the alphabet
$A^M$, that is, the new alphabet $A^M$ is the set of all $M$-letter words over $A$. 
Applying Theorem 1 to this scheme, we obtain 
\begin {theorem}\label {c3}
Let there be a source generating sequence   $ x = x_1 ... x_N $  of letters  from the alphabet $ A = \{a_0, ..., a_{L-1} \} $ with the probability distribution $ p $. If the storage system  $ \hat{\Sigma}^{N,m}  $ is applied  to the sequence $x$, then
the average  time  of  decoding of one letter ($T$) is proportional to $\log m ( \log N+\log \log L) $ and the average memory space ($S$) is not grater than $  N (\, h_M(p) + \frac{1}{M} (   \log M + \log \log  L   + O(1)) )$ bits,
where  $M=N/m$, $h_M(p) = \frac{- 1}{M} \sum_{u \in A^{M }} p(u) \log p(u)$.
\end{theorem}
For example, suppose, $m = N/\log N$. From Theorem 3 we obtain that $T$ is proportional to $ (\log N - \log \log N)  $ $(\log N + \log \log L +O(1))$ $= \log^2N (1+o(1) )$ and $ S = N (h_M(p) + \frac{1}{ \log N } (  \log \log N + \log \log L ) +o(1) \, ))$.

We can see that these two versions and  their combinations can produce many different trade-offs between time and memory, which can be useful in practice. 

\section*{Acknowledgment}
Research  was supported  by  Russian Foundation for Basic Research (grant no. 19-47-540001).


\begin{thebibliography}{10}

\bibitem{directec}
Brisaboa, N.R., Ladra, S. and Navarro, G.  DACs: Bringing direct access to variable-length codes. Information Processing and Management, 49(1), 2013, pp.392-404.


\bibitem{directec-modern}
Gilad Barucha, Shmuel T. Kleina, Dana Shapira.
A space efficient direct access data structure.
Journal of Discrete Algorithms 43 (2017) 26–37.

\bibitem{ku}
M.O.Külekci, Enhanced variable-length codes: improved compression with efficient random access, in: Proc. Data Compression Conference, DCC-2014, Snowbird, Utah, 2014, pp. 362–371.

\bibitem{squ}
Sadakane, K. and Grossi, R., 2006, January. Squeezing succinct data structures into entropy bounds. In SODA (Vol. 6, pp. 1230-1239).

\bibitem{bm-}
J. Barbay, M. He, J. I. Munro, and S. S. Rao. Succinct indexes for strings, binary
relations and multi-labeled trees. In Proc. 18th SODA, pages 680–689, 2007.

\bibitem{ne}
J. Barbay, T. Gagie, G. Navarro, and Y. Nekrich.
Alphabet Partitioning for Compressed
Rank/Select and Applications. In International Symposium on Algorithms and Computation, 2010,  pp. 315-326. 

\bibitem{fm}
P. Ferragina, G. Manzini, Indexing compressed text, Journal of the ACM 52 (4) (2005) 552–581.


\bibitem{ks}
Klein ST, Shapira D. Random access to Fibonacci encoded files. Discrete Applied Mathematics. 2016 Oct 30;212:115-28.

\bibitem{20}
Pibiri GE, Venturini R. Techniques for inverted index compression. ACM Computing Surveys (CSUR). 2020 Dec 6;53(6):1-36.



\bibitem{bil}
Philip Bille,  Mikko Berggren Ettienne,  Inge Li G{\o}rtz and
               Hjalte Wedel Vildh{\o}j,
 Time-Space Trade-Offs for Lempel-Ziv Compressed Indexing,
 http://arxiv.org/,  2018.  




\bibitem{co}
T.~M. Cover and J.~A. Thomas, \emph{Elements of information theory}.\hskip 1em
  plus 0.5em minus 0.4em\relax New York, NY, USA: Wiley-Interscience, 2006.

\bibitem{limh}
Larmore LL, Hirschberg DS. A fast algorithm for optimal length-limited Huffman codes. Journal of the ACM (JACM). 1990 Jul 1;37(3):464-73.

\bibitem{limh2}
Baer MB. D-ary bounded-length Huffman coding. In2007 IEEE International Symposium on Information Theory 2007 Jun 24 (pp. 896-900). IEEE.

\bibitem{rae}
Boris Ryabko, Jaakko Astola, Karen Egiazarian. Fast Codes for Large Alphabets.
  Communications in Information and Systems, 2003, v.3, n. 2, pp.139-152. 
  
 \bibitem{rr}
  Boris Ryabko, Jorma Rissanen. Fast Adaptive Arithmetic Code for Large Alphabet Sources with Asymmetrical Distributions.
  IEEE Communications Letters,v. 7, no. 1, 2003,pp.33- 35.



\bibitem{fastd}
Walder J, Kratky M, Bača R, Platos J, Snasel V. Fast decoding algorithms for variable-lengths codes. Information sciences. 2012 Jan 15;183(1):66-91.

\bibitem{r1}
Ryabko B. A fast on-line code.  \emph{ 
Dokl. Akad. Nauk SSSR }. 1989, 306 (3):,548-552; (in Russian),
  translation in English  \emph{  Soviet Math. Dokl.} 1989), no.(3): 533--537.


\bibitem{r2}
Ryabko B.  A fast on-line adaptive code,    
 \emph{  IEEE Transactions on Information Theory}. 1992; 28(4): 1400 - 1404.



  


\end{thebibliography}

\end{document}